\documentclass[conference]{IEEEtran}

\usepackage{graphicx}
\usepackage{epstopdf}
\usepackage{amsmath}

\usepackage{algorithm2e}

\begin{document}
%
\title{A Dynamic Spectrum Access on SDR for IEEE 802.15.4 networks}


\author{
\IEEEauthorblockN{
Rafik Zitouni $^{\ddagger}$, Laurent George $^{\dagger}$ and Yacine Abouda 
}
\IEEEauthorblockA{
 $^{\ddagger}$ ECE Paris-LACSC Laboratory, $^{\ddagger}$ LISSI / UPEC \\
 $^{\dagger}$ UPEMLV, LIGM/ ESIEE Paris \\
37 Quai de Grenelle, 75015, Paris, France\\
Email:  zitouni@ece.fr, Laurent.George@univ-mlv.fr and abouda@ece.fr }
}

\maketitle

\begin{abstract}
Our paper deals with a Dynamic Spectrum Access (DSA) and its implementation on a Software Defined Radio (SDR) for IEEE 802.15.4e Networks. The network nodes select the carrier frequency after Energy-Detection based Spectrum Sensing (SS).  To ensure frequency hoping between two nodes in IEEE 802.15.4e Network, we propose a synchronization algorithm. We considerate the IEEE 802.15.4e Network is Secondary User (SU), and all other networks are Primary Users (PUs) in unlicensed 868/915 MHz and 2450 MHz bands of a Cognitive Radio (CR). However, the algorithm and the energy-sensor have been implemented over GNU Radio and Universal Software Radio Peripheral (USRP) SDR. In addition, real packet transmissions have been performed in two cases. In the first case, SU communicates in static carrier-frequency, while in the second case with the implemented DSA.  For each case, PU transmitter disturbs SU, which calculates Packet Success Rate (PSR) to measure the robustness of a used DSA. The obtained PSR is improved by 80\% when the SU accomplished DSA rather than a static access.

\end{abstract}


\IEEEpeerreviewmaketitle

\section{Introduction}

Spectrum scarcity issue in wireless communications is a main consequence of spectrum regulation and rigidity of telecommunication standards. Regulation authorities of telecommunication, such as FCC, define unlicensed spectrum bands for numerous applications in ISM bands. Potentially, IEEE 802.15.4e based Wireless Sensor Networks (WSN) uses these bands \cite{IEEE:std_12}. Under 2450 MHz band, a WSN shares the unlicensed spectrum with other networks such as Wifi (IEEE 802.11), Bluetooth (IEEE 802.15.1) and Microwave oven. However, 868/915 MHz frequency band is an alternative band that depends on geographic region, 868 MHz for Europe and 915 MHz for North America.

The crowded state of 2450 MHz band can be addressed with Dynamic Spectrum Access (DSA) or Dynamic Spectrum Sharing (DSS). Instead of static spectrum access, spectrum users can adjust the carrier frequency dynamically.  An open sharing model (or a spectrum commons) is one DSA model that deals with unlicensed bands \cite{Zhao07}. It considers interfering peers of users within a given band as a problem of medium access control. However, two classes of this model have been recapitulated in \cite{Zhao07}: central and distributed model. In our work, we are interested in distributed and cooperative models.  

Cognitive Radio (CR) is a system which senses its electromagnetic environment and dynamically adjusts its radio parameters to improve radio performances. The carrier frequency is one main parameter, adjusted by CR in order to avoid interference and efficiently to use a spectrum. DSA can be considered as a sub-system of CR, it deals with spectrum access. Similarly, Spectrum Sensing (SS) is a DSA sub-system. It provides information about a spectrum state. For example, signal strength (power) for each carrier frequency is an information returned by energy-sensor based on spectrum sensing. Using this information, DSA tries to share opportunistically a spectrum. Conventionally, Secondary Users (SUs) are opportunistic networks that occupy spectrum holes when Primary Users (PUs) are absent. In our work, an IEEE 802.15.4e network is considered as SU of frequency bands whereas all unlicensed networks are PUs. To build CR with its two subsystems DSA and SS, we show a solution based on a reconfigurable radio based on a flexible SDR.

Currently, nodes in WSNs cannot be SDRs due to a high-power consumption of SDRs. However, GPP based software transceiver can emulate functions of a wireless sensor node. In this case, nodes can deal with spectrum scarcity issues with CR \cite{Zahmati09}. In our paper, we focus on how to build an SDR of both DSA and SS and how to execute it on host computer. We choose GNU Radio \cite{GNURadio} and Universal Software Radio Peripheral (USRP) \cite{Ettus} SDR regarding their performances and open source properties. GNU Radio software handles transmission chains developed with flow graphs and executed on a host computer. In the literature, several transmission chains have been developed such as for IEEE 802.15.4 network and  \cite{Bloessl3} \cite{Zitouni12} \cite{Schmid06} \cite{Gahadza09} for energy-sensor.

In our SDR, we assemble a number of transmission chains, as several chains can be handled in one GNU Radio program. For SU receiver (Rx), we implement five chains. The two firsts are IEEE 802.15.4 Receivers (Rx) for two bands, the 2450 MHz and 868/815 MHz. The third and forth chains are Transmitter (Tx) and Rx of Gaussian Minimum Shift keying (GMSK) packets. Finally, the fifth chain is an energy-sensor Rx. Each chain is selected to transmit or receive information according to a synchronization algorithm. The particularity of our work is to be able to perform real wireless transmissions of packets and to deal with several transmission chains. 

The remainder of this paper is organized as follows. In Section \ref{Sec:Related_works} we outline related works dealing with  specifications and implementations of the IEEE 802.15.4e standard and the DSA techniques. In the section \ref{sec:DSA}, we describe the SDR of our DSA with energy-sensor and synchronization algorithm. Our experiments and results are detailed and discussed in Section \ref{sec:Experiments} and finally, we give some conclusions in Section \ref{sec:conclusion}.
 
\section{Related works}
\label{Sec:Related_works} 
 
\subsection{Related specifications}
IEEE 802.15.4e \cite{IEEE:std_12} published in 2012 is an enhanced version of IEEE 802.15.4 \cite{IEEE:std_06}. It defines the physical (PHY) and Medium Access Control (MAC) specifications for low-rate, low power, and low cost Personal Area Networks (PANs).

The IEEE 802.15.4 PHY layer operates in three different ISM bands. The 868 MHz band defines three communication channels available in Europe. Whereas the 915 MHz band can be divided into up to thirty channels, but it is available only in North America. The world wide available band is the 2450 MHz band with sixteen channels. The maximum data rates of the 868/915 MHz and 2450 MHz bands are respectively up to 100 kbps and 250 kbps. In addition, the standard defines twelve different physical layers according to the modulation technique.  The Direct Sequence Spread Spectrum (DSSS) operates with either Binary (BPSK) or Offset Quadrature Phase Shift Keying (O-QPSK) modulations at 868/915 MHz, and only O-QPSK at the 2450 MHz band. The main new contribution of 802.15.4e is an access mode based on Time Slotted Channel Hopping (TSCH) mode \cite{Guglielmo14}.

TSCH was introduced in order to increase network capacity, high reliability and predictable latency. It handles multi-channels based on channel (frequency) hopping. In the 2450 MHz band, the hopping among 16 channels is a function of time slots and on the number of available channels. Thus, a frequency is selected based on chosen previous channels and on the number of available channels. The channel allocation shows the possibility to dedicate different channels to each couple of wireless nodes. However, this allocation depends only on the time slots and not on the link quality. Link Quality Indicator (LQI) \cite{IEEE:std_12} \cite{IEEE:std_06} indicates an energy strength and quality of received data frames in a selected channel. Although the LQI is measured, the selected carrier frequency is predefined. In addition, changing dynamically channels in TSCH is not expected without MAC protocol coordination.



\subsection{Related implementations}
Using GNU Radio and USRP, several research works have been proposed on IEEE 802.15.4 standard. A first SDR implementation was provided in \cite{Schmid06}. It reproduces the O-QPSK layer in 2450 MHz frequency band. This SDR was validated performing real communication with Telos B motes. An extension was reported in \cite{Choong09} using USRP 2 with a multi-channel reception. In addition, the authors in \cite{Bloessl3} add five layers on O-QPSK physical layer in order to interact with Contiki OS wireless sensor networks. In \cite{Zitouni12}, BPSK layer was implemented in 868/915 MHZ frequency band. These works \cite{Schmid06}, \cite{Bloessl3} and \cite{Zitouni12} could be used to implement multi bands and multi specifications SDR. The frequency bands and standard specifications changing can be based on a specific criterion. For example, spectrum sensing can be used to formulate a criterion.

Surveys of DSA and SS techniques have respectively been proposed in \cite{Zhao07} and \cite{Yucek09}. The DSA techniques have been classified in three classes: Dynamic Exclusive Use Model, Open Sharing Model and Hierarchical Access \cite{Zhao07}. The Open Sharing Model employs open sharing among peer users as the basis for managing unlicensed spectral bands. Spectrum sensing techniques have been grouped with three main classes: Energy-detector based sensing, Cyclostationarity-Based Sensing and Matched-Filtering \cite{Yucek09}. 

Energy-detection based spectrum sensing is a simple SS to implement, the only one found on GNU Radio. It was proposed in \cite{Shea07}, based on time averaged Power Spectral Density. This detector was used in a general dynamic spectrum access in \cite{Gahadza09}. In \cite{Rashid11}, this energy detector was evaluated according to a probability of detecting wireless activity for cognitive radio. The works \cite{Gahadza09} and \cite{Rashid11} were not specified for a particular network. However, the spectrum sensing could be used by DSA with IEEE 802.15.4-based network.

\section{Dynamic Spectrum Access (DSA)}
\label{sec:DSA}

Our DSA follows an open sharing model (or spectrum commons). It is a DSA strategy where each network has equal rights in an unlicensed frequency bands \cite{Zhao07}. We consider the IEEE 802.15.4e 2450 MHz and 868/915 MHz bands, where this network is SU and other unlicensed users are PUs. For each band, IEEE 802.15.4e Tx/Rx chains are implemented with GNU Radio and can be reused as black boxes. In addition, we dedicate the spectrum sensing and the frequency selection only to the SU receiver. A spectrum sensor measures the energy (power) strength in a given frequency band, and according to a threshold, a carrier frequency is selected. Notice that, PUs could be based an Orthogonal Frequency-Division Multiplexing (OFDM) transmitter in these two bands.  

%
%


\subsection{Software Defined Radio Setting}
\label{sec:SDR_Setting}
The two components of our Software Defined Radio (SDR) are the USRP 1 front end and the GNU Radio software. The USRP 1 has been chosen regarding its less sampling rate compared to newest versions \textit{e.g.} USRP N210 \cite{Ettus} since. The sampling rates are sufficient to build an IEEE 802.15.4e communication and to experiment DSA. In addition, USRP 1 can hold two daughter boards. They contain two antennas, the first for Transmission/Reception (TX/RX) and the second for only Reception (RX). An SBX daughter board is used since it covers a large frequency band at radio front end, \textit{i.e.} from 400 MHz to 4000 MHz, the boards cover two 802.15.4 frequency bands of 868/915 MHz and 2400 MHz. In Section \ref{sec:Experiments}, SDR setup will be discussed. SDR chains are flow graphs built on GNU Radio toolkit. One flow graph represents a chain of software blocks written in C++ and connected through Python script.   

\begin{figure}
\centering
\includegraphics[width=3.4in]{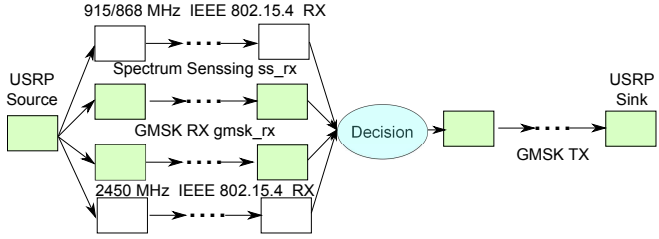}
\caption{Software chain of SU receiver (Rx)}
\label{fig:fig-rx-sdr}
\end{figure}

Tx and Rx of SU and PU are featured by a set of GNU Radio chains. Fig.\ref{fig:fig-rx-sdr} shows chains needed by a SU receiver to sense a spectrum, to coordinate a frequency selection and to receive IEEE 802.15.4 packets (data). Two receivers of 802.15.4 packets  in two frequency bands 868/915 MHz and 2450 MHz are based on \cite{Zitouni12} \cite{Schmid06}. Tx and Rx chains of GMSK packets are connected to SU receiver. In fact, through several tests, GMSK packets exchange was found reliable, \textit{i.e} every time when Tx transmits GMSK packets, Rx succeeds packet reception without a phase synchronization problem. Hence, to coordinate a frequency selection, the acknowledgment GMSK packets are exchanged. The spectrum sensing is handled by an energy-sensor chain (see Section \ref{sec:energy_sensor}). 

\begin{figure}
\centering
\includegraphics[width=3.4in]{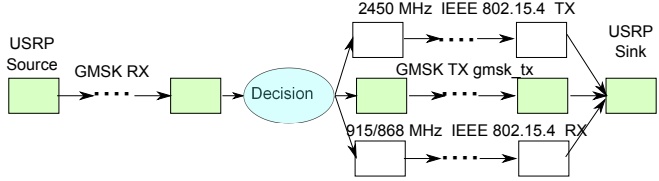}
\caption{Software chain of SU transmitter (Tx)}
\label{fig:fig-tx-sdr}
\end{figure}
\begin{figure}
\centering
\includegraphics[width=2in]{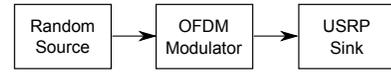}
\caption{PU transmitter (Tx)}
\label{fig:ofdm_tx}
\end{figure}

Fig.\ref{fig:fig-tx-sdr} shows SU Tx chains. Two sub transmitters are implemented for each frequency band. Similarly to the SU receiver, a frequency selection is coordinated through GMSK acknowledgment exchange. Fig.\ref{fig:ofdm_tx} highlights an SDR chain of PU Tx, which generates a random data stream and modulates it via OFDM modulator. This modulation is chosen since it is the one specified for the IEEE 802.11 standard of Wifi network. 

To separate SDR chains of SU, two daughter boards are used by the USRP module. In addition, over one daughterboard, these SDR chains can be connected to two possible antennas: Tx/Rx or Rx. For SU Rx, the GMSK Tx and Rx are carried out by the first daughter board through Tx/Rx and Rx antennas, respectively. The second daughter board supports the energy-sensor and the IEEE 802.15.4 Rx chains. Separated antennas allow the energy-sensor, GMSK Tx and Rx to be carried out continually. On the other hand, the SU Rx is similar to the USRP of SU Tx, and it contains two daughter boards, and each one supports the GMSK Tx/Rx and the IEEE 802.15.4e Rx chains.

\subsection{Energy-Sensor}
\label{sec:energy_sensor}

\begin{figure}
\centering
\includegraphics[width=3.4in]{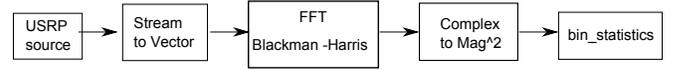}
\caption{Energy-sensor based spectrum sensing}
\label{fig:fig-ss_fg}
\end{figure}

Spectrum sensor or energy detector (see Fig.~\ref{fig:fig-ss_fg}) estimates the output of a time-averaged Power Spectral Density (PSD). For this purpose, the flow graph starts by receiving the baseband stream from USRP source.  The stream is adapted to the capacity of the USB host. Since this stream is continuous, Stream to Vector block packs a group of samples to form vectors of complex samples. Then under a Fast Fourier Transform (FFT) block, a Blackman-Harris window is useful for single tone measurement, and it is applied to each 512 sample vector. In the next block, the modulus squared is calculated averaging the magnitudes of each bin (carrier frequency) over many samples. The last block $bin\_statistics$ deals with USRP 1 constraints. 

The average energy at a given carrier frequency is calculated using the following model:

\begin{equation}
E=\frac{1}{2N}\left[ \sum_{n=-N}^{N}{\vert s(n)\vert^{2}}  \right]
\label{eq:Energy_Model}
\end{equation}

where  $N$ is the number of samples and $s(n)$ is the sample number $n$. 

In fact, an RF bandwidth from and to host computer is limited regarding USB 2 capacity limited to 8 MHz. Consequently, the frequency bandwidth to examine is divided to chunks of 8 MHz. Since a central frequency is changed via GNU Radio program, the effective change takes an extra delay on the local oscillator. During this delay or $tune\_delay$, the received samples are considered wrong and dropped. As explained in the precedent paragraph \ref{sec:SDR_Setting}, only the receiver carried out the energy-sensor.

\subsection{Dynamic frequency selection} 

The proposed algorithms \ref{Alg:rx} and \ref{Alg:tx} are message-based algorithms. They allow SU receiver and SU transmitter to decide which carrier frequency to select and how to synchronize the exchange of different packets, \textit{i.e.} the IEEE 802.15.4 and the GMSK packets.  In order to select a carrier frequency, the Rx triggers a coordination process. It starts by a spectrum sensing in a given frequency band. Then, it selects a carrier frequency which has minimum energy power. Thus, the GMSK acknowledgment messages are exchanged to ensure the effective change of the carrier frequency.

\RestyleAlgo{algoruled}
\LinesNumbered
\begin{algorithm}[ht]

  \caption{Receiver (Rx)\label{Alg:rx}}
  initialization()\;
  \While{energy $>$ threshold}{
  spectrum\_sensing(ss\_rx)\;}

  \While{not receive\_freq\_ack(gmsk\_rx)}{
  send\_new\_freq(gmsk\_tx)\;}

  \While{time $\leq$ timeout}{
  send\_clear-to-receive(gmsk\_tx)\;}

  start\_rx\_802.15.4(802\_15\_4\_rx)\;
\end{algorithm}

Algo.\ref{Alg:rx} enumerates actions of SU Rx, which senses a given frequency band and selects a carrier frequency when a sensed energy in that frequency is less than a fixed threshold (see line \textbf{(2)} to \textbf{(4)} in Algo.\ref{Alg:rx}). This threshold is taken empirically based on previous experiments. Although the energy-sensor sweeps up only to 8 MHz in one FFT window, the desired frequency band is covered by shifting this window. The energy detection is the output of the flow graph $ss\_rx$ (see Fig.\ref{fig:fig-rx-sdr} and Fig.\ref{fig:fig-ss_fg}). Thus, the new carrier frequency is selected and forwarded to 802.15.4 Tx via the $gmsk\_tx$ (see Fig.\ref{fig:fig-rx-sdr} and see also Algo.\ref{Alg:rx} from \textbf{(5)} to \textbf{(7)}). As explained above in Section \ref{sec:SDR_Setting}, one antenna is dedicated to the GMSK exchange. Since $gmsk\_rx$ demodulation is launched simultaneously with $gmsk\_tx$, the forwarding of this frequency is repeated until the reception of an acknowledgment from the SU Tx. After that, during a $timeout$, the SU Rx confirms to the SU Tx that it is clear to receive data packets (from \textbf{(8)} to \textbf{(10)} in Algo.\ref{Alg:rx}).

\begin{algorithm}[ht]
	\caption{Transmitter (Tx)} \label{Alg:tx}
 initialization()\;
 \While{not new\_freq\_received}{
	receive\_new\_frequency(gmsk\_rx)\;}
 \While{(not clear-to-receive(gmsk\_rx))
	    and (time $\leq$ timeout)}{
		send\_freq\_ack(gmsk\_tx)\;}

  \eIf{clear-to-receive(gmsk\_rx)}{
  start\_tx\_802.15.4(802\_15\_4\_tx)\;
}{receiver failed to receive clear-to-receive\;}
\end{algorithm}

The SU Tx starts data transmission only after receiving a new carrier frequency and verifying if the SU Rx is clear to receive (from \textbf{(5)} to \textbf{(7)} in Algo.\ref{Alg:tx}). An acknowledgment is transmitted using $gmsk\_tx$ to confirm the reception of a new frequency. As compared with the receiver SU Rx, the SU Tx resends acknowledgments continually during a $timeout$ until it receives a clear-to-receive message. From line \textbf{(8)} to \textbf{(12)} of Algo.\ref{Alg:rx}, the SU Tx  sends data packets only if the clear-to-receive message is received, else the reception is failed. 

\section{Experiments and results}
\label{sec:Experiments}
In our experiments, three USRP 1 devices are connected to a laptop computer in an office environment. Two devices represent SU transmitter (Tx) and receiver (Rx), whereas PU transmitter is the third one. The 868/915 MHz and 2450 MHz bands are covered by SBX daughter boards, which are plugged into a USRP 1. In GNU Radio part of the SDR, each USRP 1 is controlled via set of chains as showed in Fig.\ref{fig:fig-rx-sdr} and Fig.\ref{fig:fig-tx-sdr} of the previous Section.\ref{sec:SDR_Setting} Each chain has its parameters to initialize before and during SDR execution. 

Tab. \ref{Table:param_ss} shows offline and online parameters of spectrum sensor. The offline parameters are the sample rate and the channel bandwidth. They are initialized in the source code program before its execution. The online parameters are the bandwidth of spectrum chunks, the window's FFT, and the number of bins. They are calculated based on the offline parameters. The size of an FFT window is defined by a number of bins. It is given by Eq. \ref{Eq:fft_size}. The frequency bandwidth recovered at the software level depends on the USB port's permeability, this bandwidth is bounded bellow 8 MHz. Thus, bin\_start and bin\_stop variables are introduced to reduce the size of one FFT window by $1/8$ (see Eq.\ref{Eq:bin_start} and Eq.\ref{Eq:bin_stop}) In fact, in our experiments $80$ bins are discarded at the beginning and the end of an FFT window. Thus, the energy-sensor deals with a chunk of frequency bandwidth defined by a number of bins (or carrier frequency) spaced by a channel of 6250 Hz. For each frequency, the energy sensed is the average of the magnitudes of each bin over 512 samples (see Eq. \ref{eq:Energy_Model}) For example, the frequency band from 2405 to 2480 is divided into bandwidth chunks of 3 MHz, where the energy is calculated for each carrier frequency spaced by 6250 Hz. 

\begin{equation}
\textrm{fft\_size} = \left \lceil \frac{\textrm{usrp\_rate}}{\textrm{channel\_bandwidth}} \right \rceil
\label{Eq:fft_size}
\end{equation}

\begin{equation}
\textrm{bin\_start} = \left \lceil \frac{\textrm{fft\_size}}{8} \right \rceil 
\label{Eq:bin_start}
\end{equation}

\begin{equation}
\textrm{bin\_stop} =  \textrm{fft\_size} - \textrm{bin\_start} 
\label{Eq:bin_stop}
\end{equation}

\begin{table}[!t]
\caption{Parameters of energy-sensor} \label{Table:param_ss} \centering
\begin{tabular}{|p{1.5cm}|p{1.5cm}|p{1.5cm}|p{1.2cm}|p{1.1cm}|}
  \hline
 \textbf{USRP sample rate} & \textbf{channel bandwidth}  & \textbf{chunk of bandwidth} &\textbf{number of bins} & \textbf{FFT window} \\
 \hline 
  \hline 
 4 MS/s & 6250 Hz & 3 MHz & 480 & 640\\
   \hline 
\end{tabular}
\end{table}

\begin{figure}
\centering
\includegraphics[width=3.4in]{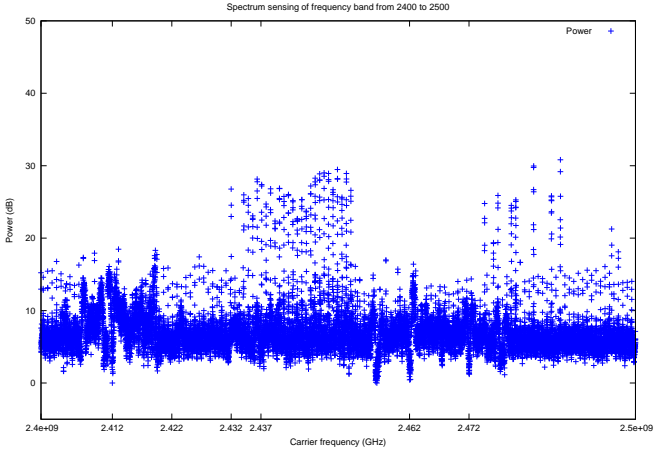}
\caption{Spectrum Sensing of frequency band 2.4 GHz to 2.5 GHz}
\label{fig:ss_2400_2500_fg}
\end{figure}

\begin{figure}
\centering
\includegraphics[width=3.4in]{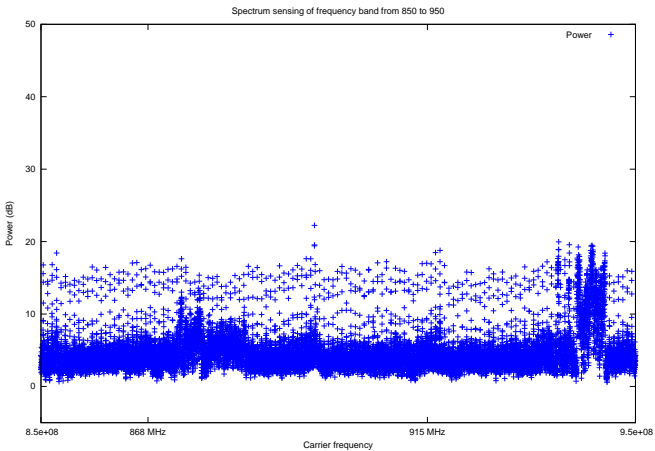}
\caption{Spectrum Sensing of frequency band 850 MHz to 950 MHz}
\label{fig:ss_850_950_fg}
\end{figure}

Since the experiments are performed in an office environment, the two targeted frequency bands of 868/915 and 2450 MHz have been sensed to get the energy power. In addition, the WiFi board of the laptop computer has detected the presence of seven IEEE 802.11 networks. Fig.\ref{fig:ss_2400_2500_fg} shows the obtained Power Spectrum Density (PSD) for each carrier frequency from 2400 MHz to 2500 MHz using the energy-sensor. Mainly, two high-power zones have been observed in the interval from 2400 MHz to 2500 MHz. In fact, the energy is up to relative power of 30 dB in intervals [2430 MHz, 2450 MHz] and [2475 MHz, 2490 MHz]. The seven detected networks have a small impact on the spectrum. In the second frequency band from 850 MHz to 950 MHz, the energy level is lower than 25 dB (see Fig.\ref{fig:ss_850_950_fg}). Hence, the detected radio-frequency activities cannot significantly disturb our experiment scenarios. 

After the characterization of the radio frequency environment, the experiments are performed through two scenarios, with and without the DSA. In the first scenario, the SUs are disturbed by OFDM transmitter, \textit{i.e} PU. The SUs exchange 7519 packets of the IEEE 802.15.4 standard, \textit{i.e} data packets, where $50$ ms is the inter-packet generation time. A disturbance is triggered at different frequencies that are around the carrier frequency of SUs. In the second scenario, the SU performs a dynamic frequency selection. Thus, the robustness of the dynamic frequency selection is measured using Packet Success Rate (PSR) and Packet Received Rate (PRR) parameters.   

\begin{figure}
\centering
\includegraphics[width=3in]{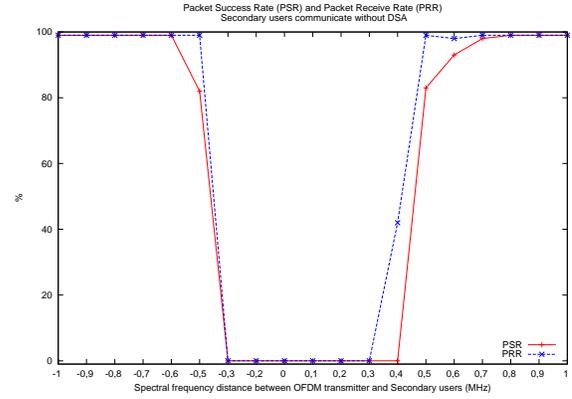}
\caption{Packet Success Rate (PSR) and Packet Received Rate (PRR) function of spectrum distance between PU and SU without DSA.}
\label{fig:psr_prr}
\end{figure}

We consider in the first scenario that the couple of SU communicates under the channel 26 (carrier frequency is 2480 MHz). The Tx generates a data stream of 1 MB and splits it into packets. Each packet has a size of 133 bytes. In addition, the data rate between Tx, and Rx is fixed to 250 kb/s (note that this rate is the same one of OQPSK PHY). Since the USRP 1 of SU transmitter and receiver are close to each other,  software amplifier $DAC$ and software gain $UHD_G$ have low values, fixed to $0.4$ and $40$ dB, respectively. For amplifying the base band signal, the constant float value $DAC$ is multiplied by the two signal components. In and Quadrature-phase. On the other hand, the $UHD_G$ is a relative gain fixed in the block of a USRP sink. However, the disturbance or the OFDM PU generates an OFDM data stream in frequencies close to that of the SU. In fact, in an interval of 2 MHz, from 2479 MHz to 2481 MHz, the OFDM signal is triggered and sweeps this interval by a step of 0.1 MHz. Fig.\ref{fig:psr_prr} shows the obtained Packet Success Rate (PSR) and Packet Received Rate (PRR) calculated using the Cyclic Redundancy Check (CRC). The PRR is calculated in the case when the packets are received but with a wrong CRC. Obviously, the PSR drop to 0 when the spectrum distance between PU and SU is lower than 0.3 MHz. Indeed, the PSR and the PRR are low since the SU Tx cannot detect the PU Tx.

The second scenario proceeds like the first one but the SU adopts DSA to avoid PU disturbance. DSA is started  by SU Rx, which senses continually frequency band from 2400 MHz to 2480 MHz and the central frequency 868 MHz. Each carrier frequency is characterized by an energy level. Thus, the selected one is that with a minimum energy level. It is communicated to SU Tx following the algorithms \ref{Alg:rx} and \ref{Alg:tx}. After that, SU Tx starts data transmission. The OFDM disturbance or PU is triggered over the selected frequency. Since the SU Rx continually senses a new bandwidth chunk of 3 MHz in 2450 MHz and 868 MHz bands and selects a new carrier frequency. In the experiment, a time period needed for every chunk is $1800$ ms. Thus, a number of data packets are dropped during spectrum sensing. 

\begin{figure}
\centering
\includegraphics[width=3in]{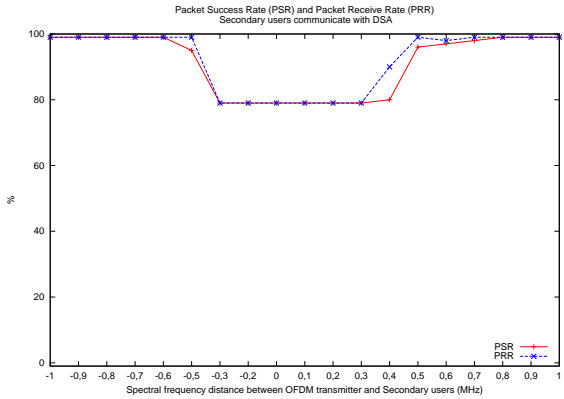}
\caption{Packet Success Rate (PSR) and Packet Received Rate (PRR) function of spectrum distance between PU and SU with DSA.}
\label{fig:psr_prr_dsa}
\end{figure}

Fig.\ref{fig:psr_prr_dsa} shows that PSR and PRR fall approximately by 20\%, when PU is at spectral distance of 0.3 MHz. In fact, this packet loss results from the extra time required for the spectrum sensing and the frequency selection. Obviously, this extra time depends on the spectrum sensing parameters (see Tab.\ref{Table:param_ss}). Using previous parameters, around $600$ ms is time to sense a band of 1 MHz. In addition, when SU Rx selects 868 MHz, the modulation change to BPSK and data rate decreases from 250 kbps to 20/40 kbps. 

With DSA, the SU improves by 80\% the PSR than a classical transmission over a static channel. This result depends on the spectrum sensor and the SU experiment parameters. In fact, this percentage can be improved if SU Tx increases inter-packet generation time, and SU Rx reduces the time to sense spectrum bandwidth. Furthermore, processing delay is introduced when we debug GNU Radio python programs by print function. 

\section{Conclusion}
\label{sec:conclusion}
In this paper, we have built dynamic spectrum access using an energy-detector based spectrum sensing. Implemented on the GNU Radio, the DSA has been performed throughout two frequency bands 868/915 MHz and 2450 MHz of the IEEE 802.15.4e standard. Communication chains of BPSK, OQPSK and energy-sensor receiver have been assembled in one SDR. To synchronize a carrier-frequency selection and to coordinate occasionally the choice of a corresponding chain, a message-based algorithm has been developed. 

Under a real packet transmission and real experimental conditions, we showed the usefulness of DSA. We improved the PSR by 80\% when we use the DSA rather than the static frequency selection, although the extra time needed for spectrum sensing and carrier frequency selection. 

Future works will focus on the implementation of this DSA on an FPGA prototype.



\end{document}